\documentstyle[editedvolume]{crckapb}

\input epsf
\def\eps@scaling{.95}
\def\epsscale#1{\gdef\eps@scaling{#1}}
\def\plotone#1{\centering \leavevmode
\epsfxsize=\eps@scaling\columnwidth \epsfbox{#1}}

\begin{opening}

\title{Cataloging of the Digitized POSS-II, 
       and Some Initial Scientific Results From It}

\author{S.G.~Djorgovski$^{1}$}
\author{R.R.~de Carvalho$^{1,2}$}
\author{R.R.~Gal$^{1}$}
\author{M.A.~Pahre$^{1}$}
\author{R.~Scaramella$^{3}$}
\author{G.~Longo$^{4}$}

\institute{$^{1}$Palomar Observatory, Caltech, Pasadena, CA 91125, USA \\
           $^{2}$Observatorio Nacional, CNPq, 20921 Rio de Janeiro, Brasil\\
           $^{3}$Osservatorio Astr.~di Roma, I-00040 Monteporzio, Italy \\
           $^{4}$Osservatorio Astr.~di Capodimonte, I-80131 Napoli, Italy}

\runningtitle{Digitized POSS-II}

\end{opening}

\begin{document}

\section{ The Survey} 

The Second Palomar Sky Survey (POSS-II) is now nearing completion.  It will
cover the entire northern sky with 894 fields ($6.5^\circ$ square) at $5^\circ$
spacings, and no gaps in the coverage.  Plates are taken in
three bands:
IIIa-J + GG395, $\lambda_{eff} \sim 480$ nm;
IIIa-F + RG610, $\lambda_{eff} \sim 650$ nm; and
IV-N   + RG9,   $\lambda_{eff} \sim 850$ nm.
Typical limiting magnitudes reached are $B_J \sim 22.5$, $R_F \sim 20.8$, and
$I_N \sim 19.5$, i.e., $\sim 1^m - 1.5^m$ deeper than the POSS-I.  The image
quality is improved relative to the POSS-I, and is comparable to the southern
photographic sky surveys.  For more details, see Reid {\it et al.} (1987), and
Reid \& Djorgovski (1993). 

The plates are being digitized at STScI, using modified PDS scanners (see the
papers by McLean, Lasker, {\it et al.} in this volume).  Plates are scanned
with 15-micron (1.0 arcsec) pixels, in rasters of 23,040 square, giving 
$\sim 1$ GB/plate, or $\sim 3$ TB of pixel data total for the entire digital
survey (DPOSS).  Preliminary astrometric solutions are good to $\sim 0.5$
arcsec, and will get better soon.  Completion of the scanning should follow
closely the completion of the plate taking, hopefully by by mid-1998.  The
plates are also digitized independently at USNOFS by Monet {\it et al.}

There is a major ongoing effort at Caltech to process and calibrate the scans, 
and catalog and classify all objects detected down to the survey limit.  We are
using SKICAT, a novel software system developed for this purpose (Weir {\it
et al.} 1993ab, 1994, 1995abc; Djorgovski {\it et al.} 1994; Fayyad {\it et al.}
1996).  SKICAT incorporates some standard astronomical image processing
packages, commercial Sybase DBMS, as well as a number of artificial
intelligence (AI) and machine learning based modules. 

As of late 1996, some 15\% of the entire survey area has been processed, but
this work will soon speed up considerably.  We have started a collaborative
effort between the Caltech DPOSS group and the Observatories of Rome and Naples
(Italy), to complete the DPOSS processing in a timely manner (project CRONA). 
Data processing pipelines are now being set up at both Italian sites.
Observatorio Nacional in Rio de Janeiro (Brasil) will join the consortium in
1997. 

The resulting Palomar-Norris Sky Catalog (PNSC) will contain all objects down
to an equivalent limiting magnitude of $B_J \sim 22^m$, with star-galaxy
classification accurate to 90\% or better down to $B_J \sim 21^m$.  The PNSC is
expected to contain $> 50$ million galaxies, and $> 2$ billion stars, including
$\sim 10^5$ quasars.  We note that the size of the DPOSS data set, in terms of
the bits, numbers of sources, and resolution elements, is $\approx 1,000
~\times$ the entire IRAS data set, and is $\approx 0.1 ~\times$ the anticipated
SDSS data set. 

We will publish the catalogs as soon as the validation tests are complete, and
the funding allows it, via computer networks and other suitable media.  The
anticipated project completion timescale is $\sim 3$ years. 

\section{ The Data} 

A particular strength of SKICAT is the star-galaxy classification, which uses
artificial induction decision tree techniques.  By using these methods, and
using superior CCD data to train the AI object classifiers, we are able to
achieve classification accuracy of 90\% or better down to $\sim 1^m$ above the
plate detection limit; traditional techniques achieve comparable accuracy
typically only $\sim 2^m$ above the detection limit.  This effectively triples
the number of usable objects for most scientific applications of these data,
since in most cases one wants either stellar objects or galaxies. 

Future technical developments include an improved treatment of very bright
and/or extended objects, optimization of the object measurement module for
crowded regions (e.g., low Galactic latitudes), better structuring of the
catalog database for efficient access and manipulation, and testing and
implementation of novel methods for data exploration, including unsupervised
classifiers and clustering analysis algorithms, etc.  Some initial results have
been presented by de Carvalho {\it et al.} (1995). 

An extensive CCD calibration effort is now underway at the Palomar 60-inch
telescope, and we expect it to expand to other sites soon.  The data are
calibrated in the Gunn $gri$ system.  We obtain at least 2 CCD fields per sky
survey field, and sometimes more.  These CCD images are used both for magnitude
zero-point calibrations, and for training of automated star-galaxy classifiers.
In addition to the CCD calibrations, we use heavily smoothed sky measurements
from the plate scans themselves (after the object removal) to ``flatfield''
away the telescope vignetting affects and the individual plate emulsion 
sensitivity variations.

As a result, we have demonstrated an unprecedented photometric stability and
accuracy for this type of photographic plate material (Weir {\it et al.}
1995a).  We have performed tests using both CCD sequences and plate overlaps,
and find that our magnitude zero-points are stable to within a few percent,
both across the plates, both between adjacent plates, and across the individual
plates.  Typical r.m.s. in the magnitude zero-points between different plates
is in the range $0.015^m - 0.045^m$ in the $r$ band, and slightly worse in the
$g$ band, perhaps due to the larger color terms in the J/$g$ calibration.
Keeping the systematic magnitude zero-point errors below 10\% is essential for
many scientific applications of these data. 

This may be best illustrated in the internal consistency of galaxy counts from
sets of adjacent POSS-II plates.  Compare the counts published by Picard
(1991), who used Cosmos machine scans processed in a traditional way, with the
counts from Weir {\it et al.} (1995a), who used DPOSS scans processed with
SKICAT.  While Picard has seen large plate-to-plate variations (a factor of 2,
or more) in number counts at a given magnitude, Weir finds excellent agreement,
to within the Poissonian errors, and reaches a magnitude deeper; yet both used
the same kind of plate material. 

Median random magnitude errors for stellar objects in all three bands start
around $0.05^m$ at the bright end, and increase to 
$\sim 0.25^m$ at $g_{lim} \approx 22^m$,
$\sim 0.20^m$ at $r_{lim} \approx 21.5^m$, and
$\sim 0.25^m$ at $i_{lim} \approx 20^m$.
For galaxies, these errors are typically higher by about 50\% at a given
magnitude.

\section{ Some Initial Scientific Applications} 

This large new database should be a fertile ground for numerous scientific
investigations, for years to come.  The nature of the data dictates its uses:
these images are not very deep by modern standards, but they do cover a very
large solid angle, and do so uniformly.  In addition to the obvious
applications such as large-scale optical identifications of sources from other
wavelengths (e.g., radio, x-ray, IR), there are two kinds of studies which can
be pursued very effectively with data sets of this size: 

(1) Statistical astronomy studies, where the sheer large numbers of detected
sources tighten the statistical errors and allow for more model parameters to
be constrained meaningfully by the data. 

(2) Searches for rare types of objects.  For example, at intermediate Galactic
latitudes, about one in a million stellar objects down to $r \approx 19.5^m$ is
a quasar at $z > 4$, yet we can find such quasars very efficiently. 

We have already started a number of scientific projects using DPOSS, which also
serve as scientific verification tests of the data, and which
have helped us catch some errors and improve and control the data quality.

Galaxy counts and colors in 3 bands from DPOSS can serve as a baseline for
deeper galaxy counts and a consistency check for galaxy evolution models.  Our
initial results (Weir {\it et al.} 1995a) show a good agreement with simple
models of weak galaxy evolution (e.g., Koo {\it et al.} 1995) at low redshifts,
$z \sim 0.1 - 0.3$.  We are now expanding this work to a much larger area, to
average over the local large-scale structure variations.  Our galaxy catalogs
have been used as input for redshift surveys down to $\sim 21^m$, e.g., in the
Palomar-Norris survey (Small {\it et al.} 1997), and several other groups plan
to use our catalogs for their own redshift surveys. 

Galaxy $n$-point correlation functions and power spectra of galaxy clustering
provide useful constraints of the CDM and other scenarios of large scale
structure.  Our preliminary results from a limited area near the NGP (Brainerd
{\it et al.} 1995; and in prep.) indicate that there is less power at large
scales than was found by the APM group (Maddox {\it et al.} 1989) in their
southern survey.  We suspect that field-to-field magnitude zero-point
calibration errors and errors in star-galaxy separation in the APM data may
account for this discrepancy.  High quality, uniform calibrations are
absolutely essential for this task.  We are now also starting to explore the
correlations of our galaxy counts with H I, IRAS, and DIRBE maps, in order to
better quantify the foreground Galactic extinction.  We expect that we can
generate extinction maps superior to those now commonly used. 

We are now starting a project to generate an objectively defined, statistically
well defined catalog of rich clusters of galaxies.  We estimate that eventually
we will have a catalog of as many as 20,000 rich clusters of galaxies at high
Galactic latitudes in the northern sky.  Their median redshift is estimated to
be $\langle z \rangle \sim 0.2$, and perhaps reaching as high as $z \sim 0.5$. 

There are many cosmological uses for rich clusters of galaxies.  They provide
useful constraints for theories of large-scale structure formation and
evolution, and represent valuable (possibly coeval) samples of galaxies to
study their evolution in dense environments.  Studies of the cluster two-point
correlation function are a powerful probe of large-scale structure, and the
scenarios of its formation.  Correlations between optically and x-ray selected
clusters are also of considerable scientific interest.  Most of the studies to
date have been limited by the statistical quality of the available cluster
samples.  For instance, the subjective nature of the Abell catalog has been
widely recognized as its major limitation.  Still, many far-reaching
cosmological conclusions have been drawn from it.  There is thus a real need to
generate well-defined, objective catalogs of galaxy clusters and groups, with
well understood selection criteria and completeness. 

We use only objects classified as galaxies in DPOSS catalogs, down to $r =
19.6^m$, where the accuracy of object classifications is $> 90$\%.  We then use
colors for selection of the candidate cluster galaxies: early-type galaxies
should better delineate high-density regions.  We then use the adaptive kernel
method to create the surface density maps.  Its major advantage is that it uses
a two-step process which smooths well the low density regions, and at the same
time leaves the high density peaks nearly unaffected. Next we evaluate the
statistical significance of the density peaks using a bootstrap technique.
Typically we set our threshold at a 4.5-$\sigma$ level, where we successfully
recover all of the known Abell clusters of richness class 0 and higher, and
also find a large number of new cluster candidates which were apparently missed
by Abell.  We have also started spectroscopic follow-up of our cluster
candidates at the Palomar 200-inch telescope. 

Another ongoing project is a survey for luminous quasars at $z > 4$.  Quasars
at $z > 4$ are valuable probes of the early universe, galaxy formation, and the
physics and evolution of the intergalactic medium at large redshifts.  The
continuum drop across the Ly$\alpha$ line gives these objects a distinctive
color signature:  extremely red in $(g-r)$, yet blue in $(r-i)$, thus standing
away from the stellar sequence in the color space.  Traditionally, the major
contaminant in this type of work are red galaxies.  Our superior star-galaxy
classification leads to a manageable number of color-selected candidates, and
an efficient spectroscopic follow-up.  As of late 1996, over 25 new $z > 4$
quasars have been discovered.  We make them available to other astronomers for
their studies as soon as the data are reduced. 

Our initial results (Kennefick {\it et al.} 1995ab) are the best estimates to
date of the bright end of the quasar luminosity function (QLF) at $z > 4$, and
are in excellent agreement with the fainter QLF evaluated by Schmidt {\it et
al.} (1995).  We have thus confirmed the decline in the comoving number density
of bright quasars at $z > 4$.  There are also some intriguing hints of possible
primordial large-scale structure as marked by these quasars.  However, much
more data is needed to check this result. 

We have also stared optical identifications of thousands of radio sources,
e.g., the VLA FIRST sources (Becker {\it et al.} 1995).  Our preliminary
results indicate that there are $\sim 400$ compact radio source ID's per DPOSS
field, and we expect a comparable number of resolved source ID's.  Among the
first 10 red stellar-like ID's we have observed spectroscopically in May 1996,
we discovered a quasar at $z = 4.36$, VF 141045+340909.  We estimate that a few
tens of $z > 4$ quasars will be found in the course of this work.  Eventually,
we expect to have $> 10^5$ ID's for the VLA FIRST sources, plus many more from
other surveys. 

In the area of statistical gravitational lensing studies, we have explored the
possibility of microlensing of quasars, by looking for a possible excess of
foreground galaxies near lines of sight to apparently bright, high-$z$ QSOs
from flux-limited samples (Barton {\it et al.}, in prep.).  We find at most a
modest excess, roughly as expected from theory, in contrast to some previous
claims which used similar data (e.g., Webster {\it et al.} 1988).  We are also
planning to use our galaxy counts to explore the possible lensing magnification
of background AGN by foreground large scale structure, as proposed, e.g., by
Bartelmann \& Schneider (1994). 

Other extragalactic projects now planned include a catalog of $\sim 10^5$
brightest galaxies in the northern sky, with a quantitative surface photometry
and morphological information, automated searches for low surface brightness
galaxies, an archival search for supernov\ae\ from plate overlaps, derivation
of photometric redshift estimators for galaxies, automated optical 
identifications of IR and x-ray sources, and so on. 

Galactic astronomy should not be neglected.  Star counts as a function of
magnitudes, colors, position, and eventually proper motions as well, fitted
over the entire northern sky at once, would provide unprecedented
discrimination between different Galactic structure models, and constraints on
their parameters.  With $\sim 2 \times 10^9$ stars, such studies would present
a major advance over similar efforts done in the past.

We can also search for stars with unusual colors or variability.  We have
started a search for stars at the bottom of the main sequence and field brown
dwarf candidates, using colors: anything with $(r-i) > 2.5$ should be
interesting. At high Galactic latitudes, about one star in a million is that
red, down to the conservative limit used so far ($r < 19.5^m$).  Such a survey
can be made much more powerful with the addition of IR data. 

The same techniques we use to search for galaxy clusters can then be applied to
our star catalogs, in an objective and automated search for sparse globulars in
the Galactic halo, tidal disruption tails of former clusters, and possibly even
new dwarf spheroidals in the Local Group (recall the Sextans dwarf, found using
similar data by Irwin {\it et al.} 1990). 

\section{ Concluding Remarks} 

These, and other studies now started or planned, should produce many interesting
and useful new results in the years to come.  Availability of large data sets
such as DPOSS over the Net or through other suitable mechanisms would also
enable astronomers and their students anywhere, even if they are far from the
major research centers or without an access to large telescopes, to do some
first-rate observational science.  This new abundance of good data may 
profoundly change the sociology of astronomy. 

Nor should we discount serendipity: With a data set as large as DPOSS, there is
even an exciting possibility of discovering some heretofore unknown types of
objects or phenomena, whose rarity would have made them escape the astronomers'
notice so far.

This is a foretaste of things to come:  with DPOSS, GSC-II, and surveys to
follow (e.g., SDSS, 2MASS, etc.), we are changing the very concept of an
astronomical catalog, into a living, permanently evolving data set, which must
come along with adequate tools for its exploration.  We will need to learn new
skills, develop new data mining and exploration tools (including AI and machine
learning techniques), new data structuring paradigms and standards, and above
all, learn to ask new kinds of astronomical questions. 

\smallskip
We are grateful to our many collaborators, including Nick Weir, Joe Roden,
Julia Kennefick, Tereasa Brainerd, Usama Fayyad, Jeremy Darling, Vandana Desai,
Emil Kartalov, Paul Stolorz, Alex Gray, Daniel Stern, and Isobel Hook; to Neill
Reid, Jean Mueller, and others in the POSS-II survey team; to Barry Lasker,
Brian McLean, and others in the digitization team at STScI; and to Massimo
Capaccioli, Roberto Buonanno, and other CRONA-ies.  This work was supported at
Caltech by the NSF PYI award AST-9157412, grants from NASA, the Bressler
Foundation, and Palomar Observatory.  This paper is the CRONA Contribution
No.~1.

\end{document}